\documentclass[%
reprint,
amsmath,amssymb,
aps,
pra,
]{revtex4-1}

\usepackage{graphicx,wrapfig,lipsum}
\usepackage{subeqnarray}
\usepackage{url,hyperref} 
\usepackage{color}
\usepackage{ulem}
\usepackage{verbatim}
\usepackage{float}
\usepackage{natbib}

\newcommand{\wt}[1]{\widetilde{#1}}

\begin{document}
\title{Current-Conserving Relativistic Linear Response for Collisional Plasmas}
\author{Martin Formanek}
\email{martinformanek@email.arizona.edu}
\author{Christopher Grayson}
\email{chrisgray1044@email.arizona.edu}
\author{Johann Rafelski}
\email{johannr@arizona.edu}
\affiliation{Department of Physics,
The University of Arizona,
Tucson, AZ, 85721, USA}
\author{Berndt M\"uller}
\email{mueller@phy.duke.edu}
\affiliation{Department of Physics,
Duke University,
Durham, NC 27708-0305, USA}

\date{\today}
\begin{abstract}
We investigate the response of a relativistic plasma to electromagnetic fields in the framework of the Boltzmann equation incorporating a collision term in the relaxation rate approximation selected in a form assuring current conservation. We obtain an explicit solution for the linearized perturbation of the Fermi-Dirac equilibrium distribution in terms of the average relaxation rate $\kappa$. We study the resulting covariant, gauge invariant, and current conserving form of the polarization tensor in the ultrarelativistic and non-relativistic limits. We evaluate the susceptibility in the ultrarelativistic limit and explore their dependence on $\kappa$. Finally, we study the dispersion relations for the longitudinal and transverse poles of the propagator. We show that for $\kappa> 2\omega_p$, where $\omega_p$ is the plasma frequency, the plasma wave modes are overdamped. In the opposite case, $\kappa \ll \omega_p$, the propagating plasma modes are weakly damped.
\end{abstract}
\maketitle


\section{Introduction}

The response of relativistic plasmas to applied electromagnetic fields is of interest in many fields of physics~\cite{NAS_HED:2003,DOE-BRN:2009}, including astrophysics, intense laser interactions with matter, and relativistic heavy ion collisions. We develop and present a novel method of obtaining and studying exact solutions for the covariant polarization tensor in linear response limit. We incorporate  the nonperturbative effect of collisional interaction  in terms of relativistically covariant and invariant generalization of the average relaxation rate method. Our linear response polarization tensor satisfies current conservation requirements and is gauge invariant. In a schematic exploration of the effect of the strength of collisional interaction we demonstrate the physical relevance of these novel theoretical developments.

Several methods have been introduced to study the linear response of a collisionless ultrarelativistic plasma following the seminal work by Weldon~\cite{Weldon:1982aq}. Of particular importance to our effort are approaches using semiclassical transport theory based on the Boltzmann equation~\cite{Mrowczynski:1987jr,Mrowczynski:1989np,Blaizot:1993zk,Kelly:1994ig,Kelly:1994dh}. However,  applications of this formalism have focused on dilute plasmas where collisions can be neglected~\cite{Blaizot:2001nr}.

The effects of the collision term were mainly studied for the purpose of deriving transport coefficients, such as the electrical conductivity, of interest is study of plasma response to long-wavelength perturbations~\cite{Mrowczynski:1988xu,Heiselberg:1993cr,Ahonen:1996nq,Baym:1997gq,Ahonen:1998iz}. Transport coefficients have also been calculated in quantum field theory using effective propagators that resum thermal modifications in order to avoid infrared divergences~\cite{Heiselberg:1994ms,Arnold:2002zm,Arnold:2003zc}.
	
In prior explorations of the  plasma interaction the linear response of the plasma to an applied field is obtained while also linearizing the collision term and studying the relaxation back to the equilibrium distribution. This can be done in the  schematic relaxation time approximation where an average relaxation time $\tau$ is introduced~\cite{Mrowczynski:1988xu,Satow:2014lia} or by calculating the momentum dependent relaxation rate $\kappa(p)$ with the input of perturbative matrix elements~\cite{Ahonen:1996nq}. Here, we use the average relaxation time approximation with momentum averaged $\kappa$ to make all calculations analytically tractable.
	
The linearized relativistic collision term proposed by Anderson and Witting~\cite{Anderson:1974} takes the form
\begin{equation}\label{eq:lincoll}
C[f] = (p^\mu u_\mu) \kappa [ f_\mathrm{eq}(p) - f(x,p) ] \,,
\end{equation}
where $\kappa=1/\tau$ is the average relaxation rate, $f(x,p)$ is the phase space distribution of charged particles in the plasma, $f_\mathrm{eq}(p)$ is their equilibrium distribution and $u_\mu$ is the four-velocity of the plasma rest frame. However, the interaction in equation \,(\ref{eq:lincoll}) violates current conservation. Of course, the full collision term respects all conservation laws; thus, this violation is an artifact of the linear approximation. In order to cure this deficiency, Bhatnagar, Gross and Krook (BGK) introduced a modification of the linearized collision term for nonrelativistic plasmas~\cite{Bhatnagar:1954zz}. In this work we generalize the BGK modification of the linearized collision term to relativistic plasmas using the Anderson-Witting form Eq.\,(\ref{eq:lincoll}) and show that the resulting linear response functions satisfy current conservation and gauge invariance constraints. 

Collective oscillations in an ultra-relativistic plasma have been explored previously in linear response formulation including binary collisions with the BGK correction in two earlier publications~\cite{Carrington:2003je,Schenke:2006xu}. While Carrington {\it et al.} \cite{Carrington:2003je} studied only collective modes in a massless equilibrium plasma, Schenke {\it et al.} \cite{Schenke:2006xu} also calculated the collective plasma response in the presence of non-equilibrium particle distributions as they arise in a rapidly expanding plasma. Both of these works were focused on the internal dynamics of the plasma, for which its quasiparticle modes are of primary interest. Our work is focused on the response of a relativistic plasma to time- and space-dependent {\it external} electromagnetic fields. 

In line with this motivation, we explore the frequency and wavelength dependence of the plasma response away from its eigenmodes, in particular, the electric and magnetic susceptibilities and electric conductivity. Also, in contrast to these previous treatments which used the plasma rest frame as their starting point, our formulation is manifestly covariant and independent of the observer's reference frame. Besides clarifying certain aspects, such as the relativistic invariance of the collision term, our treatment may be more convenient as starting point for further generalizations.

We note that the relativistic Anderson-Witting form of the scattering term with BGK modification, which ensures current conservation, does not in general assure that collisions preserve energy and momentum. In a recent development Rocha, Denicol, and Noronha~\cite{Rocha:2021zcw} aimed to ensure current and four-momentum conservation in the calculation of transport coefficients for a relativistic plasma using the Chapman-Enskog method. 
	
In our derivation of the polarization tensor we follow the ideas and methods of Blaizot and Iancu~\cite{Blaizot:2001nr} for the collisionless relativistic plasma. We avoid some of the subtleties of the Blaizot-Iancu approach by computing the response functions directly in momentum space, where the equations are of algebraic rather than integro-differential type. This allows us to treat the current-conserving modification of the collision term exactly. We also show how the response functions for a collisionless plasma are recovered in the limit $\kappa \to 0$. Finally, we calculate the plasma normal modes including their damping rates.

For the sake of simplicity, our treatment here focuses on the relativistic electron-positron plasma, but it can be easily generalized to other relativistic plasmas, such as a weakly coupled quark-gluon plasma where it applies to the plasma response to both, electromagnetic fields and color fields. Our treatment can be also easily generalized to a momentum dependent relaxation rate $\kappa(p)$ at the expense of requiring numerical evaluation of the momentum integrals.
\section{Covariant kinetic theory}\label{sec:covkinetic}

The covariant Boltzmann equation reads
\begin{multline}\label{eq:boltzmanncov}
(p \cdot \partial) f(x,p) + q F^{\mu\nu} p_\nu \frac{\partial f(x,p)}{\partial p^\mu}\\ = (p\cdot u)\kappa\left[f_\mathrm{eq}(p)\frac{n(x)}{n_\mathrm{eq}} - f(x,p)\right]\,,
\end{multline}
where on the right hand side we included a collision term adopted from~\cite{Bhatnagar:1954zz}. The meaning of the 4D scalar product dot notation is given by
\begin{equation}
p \cdot u \equiv p^\mu u_\mu = \eta_{\mu\nu}p^\mu u^\nu
\end{equation}
with the flat space-time metric $\eta_{\mu\nu}$ with signature (+,-,-,-). The fluctuating and equilibrium densities are defined covariantly as
\begin{align}
\label{eq:ndef1}n(x) &\equiv 2 \int (dp)(p\cdot u)f(x,p)\,,\\
\label{eq:ndef2}n_\mathrm{eq} &\equiv 2\int (dp)(p\cdot u) f_\mathrm{eq}(p)\,.
\end{align}
The factor two accounts for the spin degrees of freedom. The covariant integration measure is defined as
\begin{equation}\label{eq:measure}
(dp) \equiv \frac{d^4p}{(2\pi)^4}4\pi \delta_+(p^2-m^2) = \left.\frac{d^3p}{(2\pi)^3p^0}\right|_{p^0 = \sqrt{|\pmb{p}|^2 + m^2}} \,,
\end{equation}
where we consider only the positive energy particles. The covariant Fermi-Dirac distribution function reads~\cite{Becattini:2013fla}:
\begin{equation}\label{eq:equilibriumFD}
f_\mathrm{eq}(p) \equiv \frac{1}{\exp(p \cdot u/T) + 1}\,.
\end{equation}
where $u^\mu$ is the 4-velocity of medium with respect to which the invariant $p \cdot u$ is measured. $T$ denotes the temperature in the medium rest frame. 

The reason why we are implementing the collision term in the form (\ref{eq:boltzmanncov}) is that this formulation explicitly conserves the 4-current
\begin{equation}\label{eq:jmudef}
j^\mu = 2q \int (dp)p^\mu f(x,p)\,,
\end{equation}
where the factor of two again accounts for the spin degrees of freedom. We can prove this statement by applying $\partial_\mu$ on this expression and substituting back from the Boltzmann equation (\ref{eq:boltzmanncov})
\begin{multline}
\partial_\mu j^\mu = 2q \int (dp) \left\{-q F^{\mu\nu}p_\nu \frac{\partial f(x,p)}{\partial p^\mu}\right. \\
\left. + (p\cdot u)\kappa \left[f_\mathrm{eq}(p) \frac{n(x)}{n_\mathrm{eq}}-f(x,p) \right] \right\}\,.
\end{multline} 
The first term should naturally vanish because the collisionless Vlasov equation preserves 4-current. This can be seen upon integration by parts and use of the antisymmetry of $F^{\mu\nu}$. On the other hand, the collision term vanishes by design - see definitions (\ref{eq:ndef1},\ref{eq:ndef2}).

We are writing the solution for the distribution function as
\begin{equation}\label{eq:perturbation}
f(x,p) = f_\mathrm{eq}(p) + \delta f(x,p)\,,
\end{equation}
where the perturbation from the equilibrium comes because of the inclusion of the external EM field. To first order in the fields we can write
\begin{multline}\label{eq:boltzdeltaf}
(p \cdot \partial) \delta f(x,p) + q F^{\mu\nu}p_\nu \frac{\partial f_\mathrm{eq}(p)}{\partial p^\mu}\\
 = (p\cdot u) \kappa \left[\frac{f_\mathrm{eq}(p)}{n_\mathrm{eq}}\delta n(x)-\delta f(x,p) \right]\,,
\end{multline}
where the quantity $\delta n(x)$ is defined following the definitions (\ref{eq:ndef1},\ref{eq:ndef2}) as
\begin{equation}
\delta n (x) \equiv 2 \int (dp) (p\cdot u)\delta f(x,p)\,.
\end{equation}
An explicit solution of the Boltzmann equation can be obtained more easily in momentum space after a Fourier transformation. We define the Fourier transform $\widetilde{g}(k^\mu)$ of a general function $g(x^\mu)$ of space-time coordinates as 
\begin{equation}\label{eq:ftdef}
\wt{g}(k) \equiv \frac{1}{(2\pi)^2} \int d^4x e^{i k \cdot x} g(x)\,.
\end{equation} 
The Fourier transformation replaces partial derivatives $\partial_\mu$ with the 4-momentum $k_\mu$:
\begin{equation}
\partial_\mu \rightarrow - i k_\mu \,,
\end{equation}
and the Fourier transformed Boltzmann equation reads
\begin{multline}\label{eq:boltzfourier}
-i (p \cdot k) \wt{\delta f}(k,p) + q\wt{F}^{\mu\nu}p_\nu \frac{\partial f_\mathrm{eq}(p)}{\partial p^\mu} \\
= (p\cdot u)\kappa \left[\frac{f_\mathrm{eq}(p)}{n_\mathrm{eq}}\wt{\delta n}(k) - \wt{\delta f}(k,p) \right]\,.
\end{multline}
In the following we simplify the notation of derivatives of the equilibrium function with respect to momentum as
\begin{equation}
\frac{\partial f_\mathrm{eq}(p)}{\partial p^\mu} = \frac{d f_\mathrm{eq}(p)}{d (p \cdot u)} u_\mu \equiv f'_\mathrm{eq}(p) u_\mu \,.
\end{equation}
Solving (\ref{eq:boltzfourier}) for $\wt{\delta f}(k,p)$ we have
\begin{multline}\label{eq:deltaftilde}
\wt{\delta f}(k,p) = \frac{i}{p \cdot k + i (p\cdot u) \kappa}\bigg[-q (u \cdot \wt{F} \cdot p)f'_\mathrm{eq}(p) \\
\left.+ (p\cdot u) \kappa \frac{f_\mathrm{eq}(p)}{n_\mathrm{eq}}\wt{\delta n}(k)\right]\,,
\end{multline}
which can be readily integrated over $2(dp)(p\cdot u)$ to obtain an equation for $\wt{\delta n}(k)$
\begin{equation}
\wt{\delta n}(k) = R(k) - Q(k)\wt{\delta n}(k)\,,
\end{equation}
where the integrals are defined
\begin{align}\label{eq:R}
R(k)  \equiv -2i \int (dp)(p\cdot u) \frac{q(u \cdot \wt{F} \cdot p)f'_\mathrm{eq}}{p \cdot k + i (p\cdot u)\kappa}\,,\\
\label{eq:Q}Q(k) \equiv -2i \frac{\kappa}{n_\mathrm{eq}}\int (dp)(p\cdot u)^2 \frac{f_\mathrm{eq}(p)}{p\cdot k + i(p\cdot u)\kappa}\,.
\end{align}
This way the solution for $\wt{\delta n}(k)$ in terms of the external fields is simply 
\begin{equation}
\wt{\delta n}(k) = \frac{R(k)}{1+Q(k)}\,.
\end{equation}
We can substitute this result back into (\ref{eq:deltaftilde}) to obtain an explicit expression for $\wt{\delta f}(k,p)$
\begin{multline}\label{eq:deltafsolution}
	\wt{\delta f}(k,p) = \frac{i}{p \cdot k + i (p\cdot u) \kappa}\bigg[-q (u \cdot \wt{F} \cdot p)f'_\mathrm{eq}(p) \\
	\left.+ (p\cdot u) \kappa \frac{f_\mathrm{eq}(p)}{n_\mathrm{eq}} \frac{R(k)}{1+Q(k)}\right]\,,
\end{multline}
where the right hand side contains only known quantities. 

\subsection{Induced current}

If we subtract the antiparticle distribution $\wt{f}_-$ from the particle distribution $\wt{f}_+$ obtain the total current in the momentum space
\begin{multline}\label{eq:currentdef}
\wt{j}^\mu(k) = 2 q \int (dp)p^\mu [\wt{f}_+(k,p) - \wt{f}_-(k,p)]\\
 = 4 q \int (dp)p^\mu \wt{\delta f}(k,p)\,.
\end{multline} 
Whereas the equilibrium contributions cancel because of the opposite sign of the charges of particles and antiparticles, the perturbations add up due the change in sign of the external force $qF^{\mu\nu}p_\nu$. 
Thus our final expression for the induced current is
\begin{equation}\label{eq:jmu}
\boxed{\wt{j}^\mu(k) = R^\mu(k) - \frac{R(k)}{1+Q(k)} Q^\mu(k)}
\end{equation}
where the integrals $R^\mu(k)$ and $Q^\mu(k)$ are defined analogously to (\ref{eq:R},\ref{eq:Q}) as
\begin{align}
\label{eq:Rmu}R^\mu(k)  \equiv -4q^2i \int (dp) p^\mu \frac{(u \cdot \wt{F} \cdot p)f'_\mathrm{eq}}{p \cdot k + i (p\cdot u)\kappa}\,,\\
\label{eq:Qmu}Q^\mu(k) \equiv -4qi \frac{\kappa}{n_\mathrm{eq}}\int (dp)(p\cdot u) p^\mu \frac{f_\mathrm{eq}(p)}{p\cdot k + i(p\cdot u)\kappa}\,.
\end{align} 
Note that we absorbed the factor $4q$ from the current (\ref{eq:jmu}) into the definition of these integrals.

\subsection{Current conservation}

We can show that the result (\ref{eq:jmu}) satisfies the continuity equation. In the momentum space this expression has to be orthogonal to $k_\mu$. The first term
\begin{multline}
k_\mu R^\mu = -4q^2i \int (dp) [k \cdot p + i(p\cdot u)\kappa - i(p\cdot u)\kappa)]\\
\times  \frac{(u \cdot \wt{F} \cdot p)f'_\mathrm{eq}}{p \cdot k + i (p\cdot u)\kappa} = -i2q\kappa R\,,
\end{multline}
where we added and subtracted $i (p\cdot u)\kappa$ in the numerator. The integral for which the denominator cancels vanishes. On the other hand
\begin{multline}
k_\mu Q^\mu = -4qi \frac{\kappa}{n_\mathrm{eq}}\int (dp) [k \cdot p + i(p\cdot u)\kappa - i(p\cdot u)\kappa)]\\
\times \frac{(p\cdot u)f_\mathrm{eq}(p)}{p\cdot k + i(p\cdot u)\kappa} = - 2qi\kappa(1+Q)\,,
\end{multline}
because the part where the denominator cancels integrates to $n_\mathrm{eq}$ by definition (\ref{eq:ndef2}). Substituting all into our result (\ref{eq:jmu}) we have
\begin{equation}
k \cdot \wt{j}(k) = -i2q\kappa R + \frac{R2qi\kappa}{1+Q}(1+Q)=0\,,
\end{equation}
as expected.

\subsection{Covariant polarization tensor}

We can compare our result (\ref{eq:jmu}) with the covariant formulation of the Ohm's law~\cite{Starke:2014tfa}. In the momentum space it reads
\begin{equation}\label{eq:ohm}
\wt{j}^\mu(k) = \Pi^\mu_\nu(k) \wt{A}^\nu(k)\,.
\end{equation}
The Fourier transform of the electromagnetic tensor in terms of the 4-vector potential in momentum space $\wt{A}^\mu$ is
\begin{equation}\label{eq:ftfmunu}
\wt{F}^{\mu\nu}(k) = -i k^\mu \wt{A}^\nu(k) + i k^\nu \wt{A}^\mu(k)\,,
\end{equation}
which we can substitute to the definition of $R^\mu(k)$ (\ref{eq:Rmu}) to obtain
\begin{multline}
R^\mu(k) = - 4q^2 \int (dp) f'_\mathrm{eq}(p)\\
\times \frac{(u\cdot k)p^\mu p_\nu - (k \cdot p)p^\mu u_\nu}{p\cdot k + i (p\cdot u) \kappa} \wt{A}^\nu(k)\,,
\end{multline}
from which we see that the contribution of $R^\mu$ to the polarization tensor is
\begin{multline}\label{eq:Rmunu}
R^\mu_\nu(k) \equiv - 4q^2 \int (dp) f'_\mathrm{eq}(p)\\
\times\frac{(u\cdot k)p^\mu p_\nu - (k \cdot p)p^\mu u_\nu}{p\cdot k + i (p\cdot u) \kappa}.
\end{multline}

Similarly we can address the contribution of the second term which is hidden in the $R(k)$ scalar. In terms of the 4-vector potential in the momentum space $\wt{A}^\nu$ we have
\begin{multline}
R(k) = - 2q \int (dp)(p\cdot u)f'_\mathrm{eq}(p)\\
\times\frac{(u\cdot k)p_\nu - (k \cdot p)u_\nu}{p\cdot k + i (p\cdot u) \kappa}\wt{A}^\nu(k)\,.
\end{multline}
We can identify in this expression a 4-vector $H_\nu(k)$ defined as
\begin{multline}\label{eq:Hnu}
H_\nu(k) \equiv - 2q \int (dp)(p\cdot u)f'_\mathrm{eq}(p)\\
\times\frac{(u\cdot k)p_\nu - (k \cdot p)u_\nu}{p\cdot k + i (p\cdot u) \kappa}
\end{multline}
so that the polarization tensor is given by
\begin{equation}\label{eq:pimunu}
\boxed{\Pi^\mu_\nu(k) = R^\mu_\nu(k) - \frac{Q^\mu(k) H_\nu(k)}{1+Q(k)},}
\end{equation}
where the covariant quantities $R^\mu_\nu$, $Q^\mu$, $H_\nu$, and $Q$ are given by the integrals (\ref{eq:Rmunu},\ref{eq:Qmu},\ref{eq:Hnu},\ref{eq:Q}) respectively. 

We briefly mention the properties of the polarization tensor. The induced charges have to satisfy the continuity equation; in momentum space this implies
\begin{equation}\label{eq:currentconserv}
k_\mu \wt{j}^\mu = 0 = k_\mu \Pi^\mu_\nu \wt{A}^\nu\,.
\end{equation}
Since the current also has to be gauge invariant
\begin{equation}\label{eq:transversal2}
\left.
\begin{array}{c}
\wt{j}'^\mu_\text{ind} = \Pi^\mu_\nu \wt{A}'^\nu = \Pi^\mu_\nu (\wt{A}^\nu - i k^\nu \wt{\chi}),\\
\wt{j}^\mu_\text{ind} = \Pi^\mu_\nu \wt{A}^\nu
\end{array}
\right\} \ \Rightarrow \Pi^\mu_\nu k^\nu = 0\,,
\end{equation}
where $\wt{\chi}(k)$ is a Fourier transform of an arbitrary gauge function. This implies that if we multiply the polarization tensor from the right with $k^\nu$ it should vanish. This property is apparent in the covariant form of $H_\nu(k)$ (\ref{eq:Hnu}) and $R^\mu_\nu(k)$ (\ref{eq:Rmunu}). 

\section{Polarization tensor}

In order to evaluate the integrals $Q^\mu$, $H_\nu$, $Q$ and $R^\mu_\nu$ we choose to orient the vector $\pmb{k}$ along the $z$-axis
\begin{equation}
k^\mu = (\omega,0,0,k)\,.
\end{equation}
With this choice the gauge condition imposed on the polarization tensor (\ref{eq:transversal2}) becomes
\begin{equation}\label{eq:gaugeincomponents}
\Pi^0_z = -\frac{\omega}{k}\Pi^0_0, \quad \Pi^z_z = - \frac{\omega}{k}\Pi^z_0\,.
\end{equation}
We choose to perform the calculations in the rest frame of the medium $u^\mu = (1,0,0,0)$ and in two limits where, either the mass can be neglected (ultrarelativistic limit), or where the plasma particle move slowly compared with the speed of light (nonrelativistic limit).

\subsection{Ultrarelativistic limit}

In this limit we neglect the mass of the plasma particles as compared to their momentum. With this assumption the non-zero components of the polarization tensor are (see Appendix \ref{sec:ultrarel} for details):
\begin{align}
\Pi^0_0 &= - m_D^2 L \left( 1+ \frac{i\kappa\Lambda}{2k-i\kappa\Lambda} \right)\,,\\
\Pi^0_z &=m_D^2 \frac{\omega}{k}L \left( 1+ \frac{i\kappa\Lambda}{2k-i\kappa\Lambda} \right)\,,\\
\Pi^z_0 &= -m_D^2\frac{\omega}{k}L\left(1 + \frac{i\kappa \Lambda}{2k - i\kappa \Lambda}\right)\,,\\
\Pi^z_z &= m_D^2\frac{\omega^2}{k^2}L\left(1 + \frac{i\kappa \Lambda}{2k - i\kappa \Lambda}\right)\,,\\
\Pi^x_x &= \Pi^y_y = \frac{m_D^2\omega}{4k}\left( \frac{\omega'^2}{k^2}\Lambda - \Lambda - \frac{2\omega'}{k}\right)\,.
\end{align}
The quantities $\omega'$, $\Lambda$, and $L$ are defined as
\begin{align}\label{eq:definitions}
\omega' \equiv \omega + i\kappa, \quad \Lambda \equiv \ln \frac{\omega' + k}{\omega'-k}, \quad L \equiv 1-\frac{\omega'}{2k}\Lambda\,,
\end{align}
and the Debye screening mass $m_D^2$ is given by integral (\ref{eq:mD}):
\begin{equation}\label{eq:debye}
m_D^2 \equiv \frac{q^2T^3}{3}\,.
\end{equation}

\subsection{Nonrelativistic limit}

In this limit we assume that the velocity of the plasma particles is much smaller than the speed of llight, i.~e., we work in the low temperature limit $T \ll m$. We also assume that $|\pmb{p}|k/m\omega' \ll 1$ in order to be able to perform the expansion in the momentum integrals. We refer to Appendix \ref{sec:nonrel} for the derivation of the results. The non-zero components of the polarization tensor are: 
\begin{align}
\Pi^0_0 &= m_L^2 \frac{k^2}{\omega'^2} \frac{1}{1-\frac{i\kappa}{\omega'}\left(1+\frac{T k^2}{m\omega'^2} \right)}\,,\\
\Pi^0_z &= -m_L^2 \frac{\omega k}{\omega'^2} \frac{1}{1-\frac{i\kappa}{\omega'}\left(1+\frac{T k^2}{m\omega'^2} \right)}\,,\\	
\Pi^z_0 &= m_L^2 \frac{\omega k}{\omega'^2} \frac{1}{1-\frac{i\kappa}{\omega'}\left(1+\frac{T k^2}{m\omega'^2} \right)}\,,\\	
\Pi^z_z &= -m_L^2 \frac{\omega^2}{\omega'^2} \frac{1}{1-\frac{i\kappa}{\omega'}\left(1+\frac{T k^2}{m\omega'^2} \right)}\,,\\
\Pi^x_x &= \Pi^y_y = -m_L^2 \frac{\omega}{\omega'}\,.
\end{align}
Here we have introduced the mass scale $m_L^2$ through the integral (\ref{eq:mL})
\begin{equation}
m_L^2 \equiv q^2 \left(\frac{2mT}{\pi}\right)^{3/2}\frac{e^{-m/T}}{2m}\,.
\end{equation}	
We shall see in Section \ref{sec:disp} below that $m_L$ is the nonrelativistic plasma frequency.

\subsection{Current conservation in components}

The current conservation relation (\ref{eq:currentconserv}) reads in explicit component notation:
\begin{equation}
0 = k_0 \Pi^0_0 \wt{A}^0 + k_0 \Pi^0_z \wt{A}^z + k_z \Pi^z_0 \wt{A}^0 + k_z \Pi^z_z \wt{A}^z\,.
\end{equation}
To do show that the right-hand side vanishes as required, we invoke the Lorentz gauge condition $k \cdot \wt{A}=0$ which allows us to express $\wt{A}^z$ in terms of $\wt{A}^z$:
\begin{equation}
\wt{A}^z = \frac{\omega}{k}\wt{A}^0\,.
\end{equation}
The current conservation condition now becomes a constraint on the polarization tensor alone:
\begin{equation}\label{eq:currentcomputation}
0 = \Pi^0_0 + \frac{\omega}{k}\Pi^0_z - \frac{k}{\omega}\Pi^z_0 - \Pi^z_z\,. 
\end{equation}
If we invoke the gauge invariance constraint (\ref{eq:gaugeincomponents}) we can simplify
\begin{equation}
0 = \left(1-\frac{\omega^2}{k^2} \right)\left(\Pi^0_0 - \Pi^z_0 \frac{k}{\omega}\right)\,.
\end{equation}
The current conservation condition is now reduced to the constraint
\begin{equation}\label{eq:currentconservcomp}
\Pi^z_0 = \frac{\omega}{k}\Pi^0_0\,,
\end{equation}
which is obviously satisfied in both limits. In conclusion, the polarization tensor $\Pi^\mu_\nu$ conserves the electric current and ensures its gauge invariance. With both covariant indices $\Pi_{\mu\nu}$ is even symmetric.

\section{Collective Plasma Behavior}

The equivalent dielectric and susceptibility tensors, through  which  one  can  study  collective  plasma  effects, are defined through the spatial portion of the polarization tensor $\Pi^i_j$ as in~\cite{Starke:2014tfa,Melrose:2008},
\begin{equation}\label{dielten}
     K^i_j(\omega,k) = \varepsilon^i_j/\varepsilon_0 = 1+\frac{\Pi^i_j(\omega,k)}{\omega^2} = 1+\chi^i_j(\omega,k)
\end{equation}
For an isotropic plasma the spatial part can be written in terms of two independent response functions, one ($\Pi_T$) transverse to $\mathbf{k}$, the other ($\Pi_L$) parallel to $\mathbf{k}$. These are immediately recognizable in the spatial components of $\Pi^{\mu}_{\nu}$, evaluated in the rest frame, given that $\mathbf{k}$ was chosen to be in the $\hat{z}$ direction~\cite{Melrose:2008}. The longitudinal and transverse response functions are given by
\begin{equation}\label{eq:piLT}
    \Pi_L =\Pi^z_z = - \frac{\omega^2}{k^2}\Pi^0_0, \quad \Pi_T =\Pi^x_x=\Pi^y_y \,.
\end{equation}
Both prescriptions for $\Pi_L$ are equivalent because of current conservation. They arise from solving the Maxwell equations in medium for either $A_0$ or $A_\parallel$. Reference frame independent definitions of $\Pi_L$ and $\Pi_T$ can be given with the help of manifestly Lorentz invariant projection operators~\cite{Melrose:2008}.

\subsection{Susceptibilities}

\begin{figure}[H]
\centering
\includegraphics[width=0.95\linewidth]{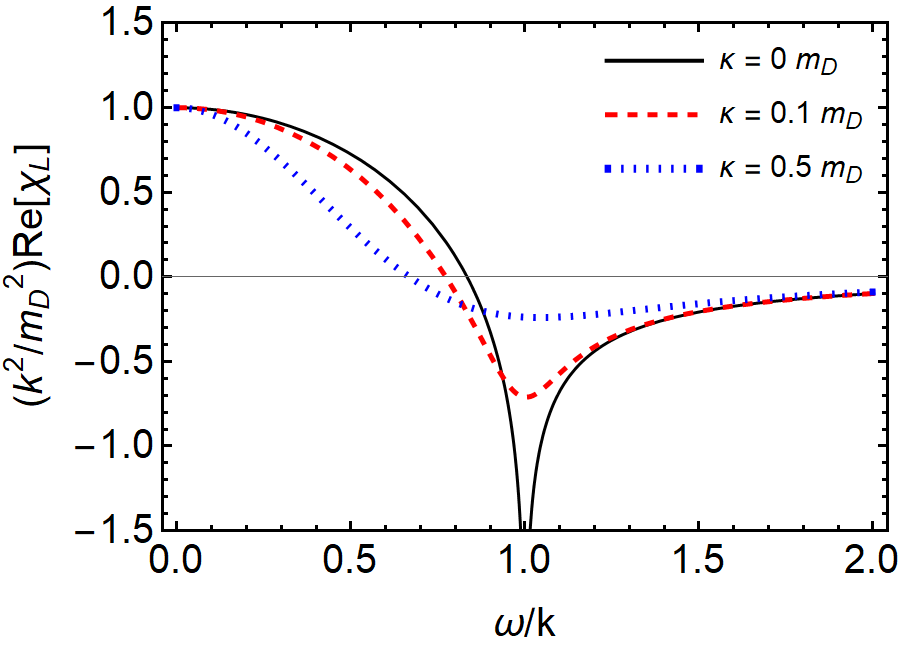}
\caption{The real part of the longitudinal susceptibility is plotted as a function of phase velocity $\omega/k$ for various values of the relaxation rate $\kappa$.The curve for $\kappa=0$ matches the electric susceptibility in~\cite{Weldon:1982aq}. The addition of the collision term smoothens the frequency dependence.}
\label{fig:Re_chi_L}
\end{figure}

\begin{figure}[H]
\centering
\includegraphics[width=0.95\linewidth]{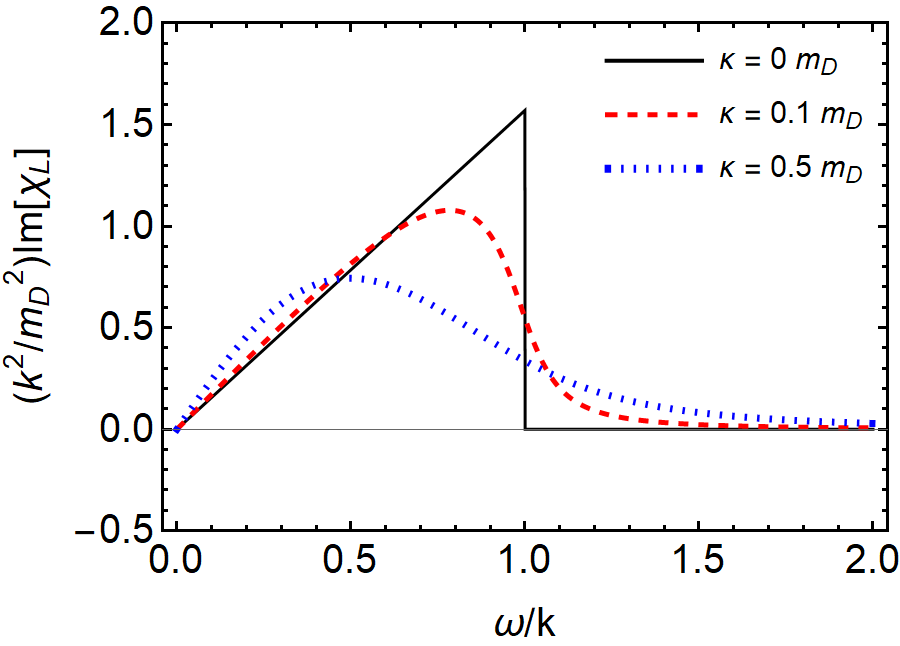}
\caption{The imaginary part of the longitudinal susceptibility is plotted as a function of phase velocity $\omega/k$ for various values of relaxation rate. The discontinuity of ${\rm Im}[\chi_L]$ at $\omega/k = 1$ is smoothened out for $\kappa > 0$.}
\label{fig:Im_chi_L}
\end{figure}

The susceptibilities describes the frequency and wavelength dependent response of the plasma to an external field. One can define the transverse and longitudinal susceptibilities through equation (\ref{dielten}) using the transverse and longitudinal response functions,
\begin{equation}\label{eq:chi}
    \chi_L =\frac{\Pi_L}{\omega^2}, \quad \chi_T = \frac{\Pi_T}{\omega^2}
\end{equation}
The real and imaginary parts of these susceptibilities are shown in Figs.~\ref{fig:Re_chi_L}--\ref{fig:chi_T3d}.
 
The solid (black) line in Fig.~\ref{fig:Re_chi_L} for $\kappa=0$ reproduces the result shown in Fig.~1 of Ref.~\cite{Weldon:1982aq}. The dashed (red) and dotted (blue) lines show how the real part longitudinal susceptibility is modified for non-zero values of the collision rate ($\kappa/m_D = 0.1, 0.5$). Figure \ref{fig:Im_chi_L} shows the imaginary part of the longitudinal susceptibility as a function of $\omega/k$. One sees how the Landau damping, which in the absence of collisions ($\kappa = 0$) is constrained to the space-like momentum domain, extends into the time-like domain for $\kappa > 0$. 

Figures~\ref{fig:chi_L3d}~and~\ref{fig:chi_T3d}~show three-dimensional plots of the real and imaginary part of $\chi_L(\omega,k)$ and $\chi_T(\omega,k)$ for $\kappa/m_D = 0.1$, respectively. The normal modes are clearly visible as features along the diagonal in the $\omega-k$ plane.
We can expand the susceptibilities in the optical limit, $k \ll m_D$. In this limit, the longitudinal and transverse susceptibilities are given by
\begin{equation}\label{chi_l_opt}
   \chi_L(\omega,k) =-\frac{\omega_p^2}{\omega(\omega+i\kappa)} - \frac{\omega_p^2}{\omega^2} \frac{ ( 9 \omega+5i\kappa )}{15(\omega+ i \kappa )^3} k^2 + O\left(k^4\right) \,,
\end{equation}
\begin{equation}
   \chi_T(\omega,k) = -\frac{\omega_p^2}{\omega(\omega+i\kappa)}-\frac{\omega_p^2 k^2}{5 \omega (\omega +i \kappa )^3} + O\left(k^4\right) \,,
\end{equation}
where $\omega_p^2 = m_D^2/3$ is the plasma frequency. Here we can clearly see that to leading order the two susceptibilities are the same as it must be for an isotropic medium. Solving the dispersion relations for the $k$ independent term gives the plasma frequency, this is done in (\ref{sec:disp}). 

We can also expand the susceptibilities in the limit $k\gg m_D$ where one finds:
\begin{equation}
     \chi_L(\omega,k) = \frac{m_D^2}{k^2}-\frac{ i \pi  \omega  m_D^2}{2 k^3}+O\left(\frac{1}{k^4}\right)
\end{equation}
\begin{multline}
    \chi_T(\omega,k) = -\frac{ i \pi m_D^2}{4 k \omega }-\frac{ m_D^2 (\omega +i \kappa )}{k^2 \omega }
    \\
    +\frac{ i \pi m_D^2 (\omega +i \kappa )^2}{4 k^3 \omega }+O\left(\frac{1}{k^4}\right)
\end{multline}
The frequency independent term in the longitudinal susceptibility describes the phenomenon of static Debye screening, with the inverse Debye mass $1/m_D$ defining the screening length. 

\begin{widetext}
\phantom{Phantom text}

\begin{figure}[H]
\centering
\includegraphics[width=0.45\linewidth]{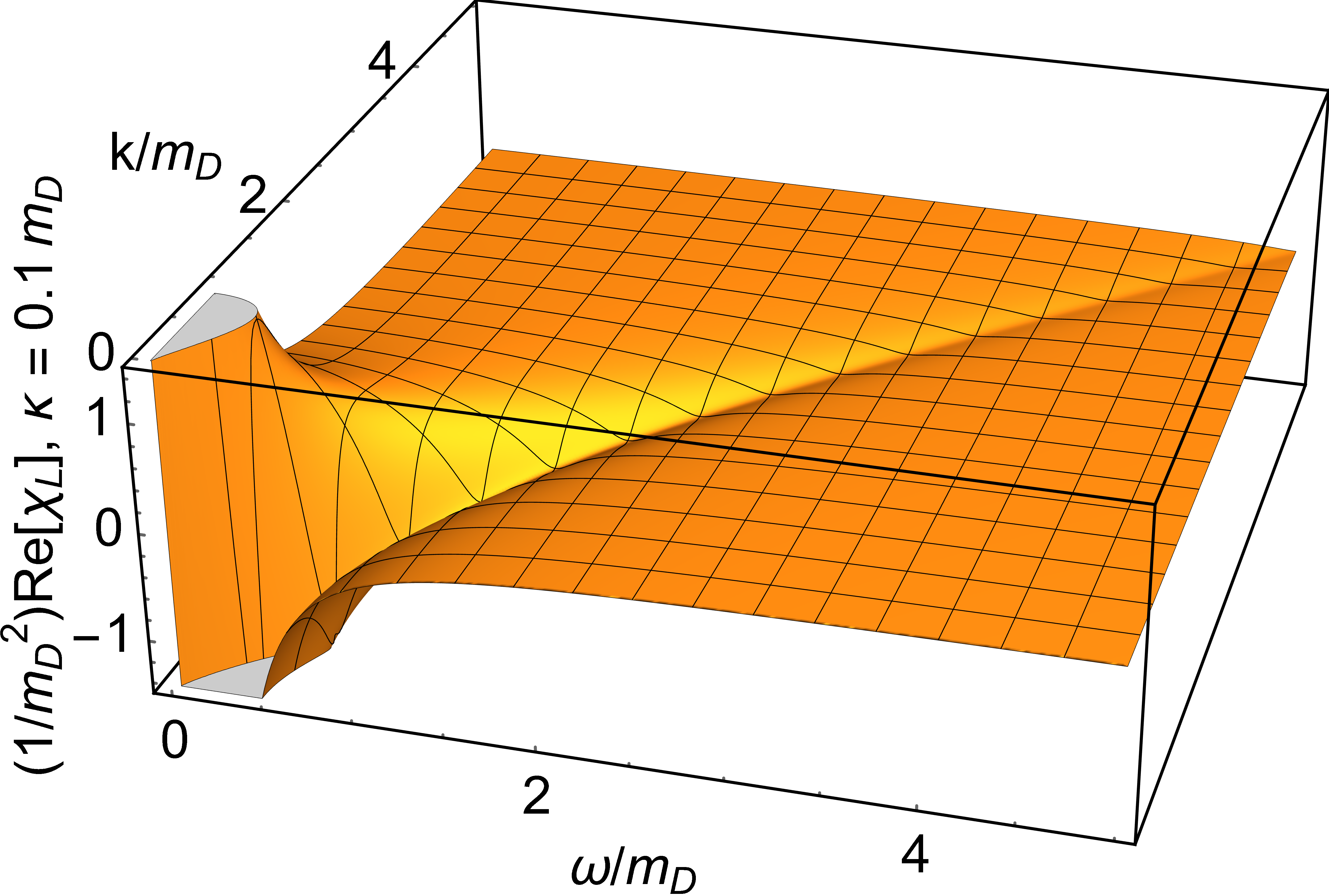}
\hspace{0.05\linewidth}
\includegraphics[width=0.45\linewidth]{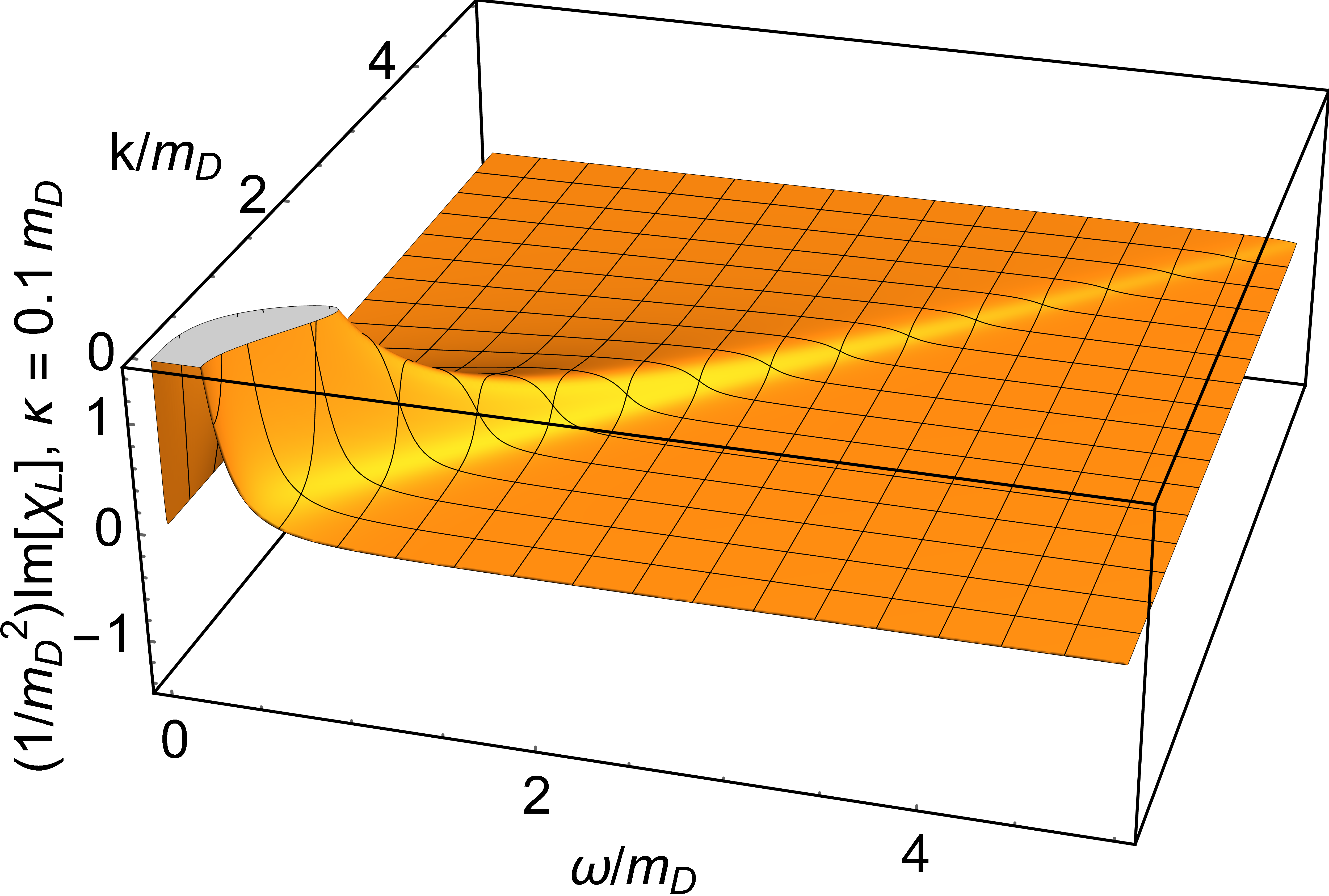}
\caption{Longitudinal susceptibility $\chi_L(\omega, k)$ in units of $m_D^2$ for $\kappa = 0.1\,m_D$. Left panel: real part of $\chi_L(\omega,k)$; right panel: imaginary part of $\chi_L(\omega,k)$.}
\label{fig:chi_L3d}
\end{figure}

\begin{figure}[H]
\centering
\includegraphics[width=0.45\linewidth]{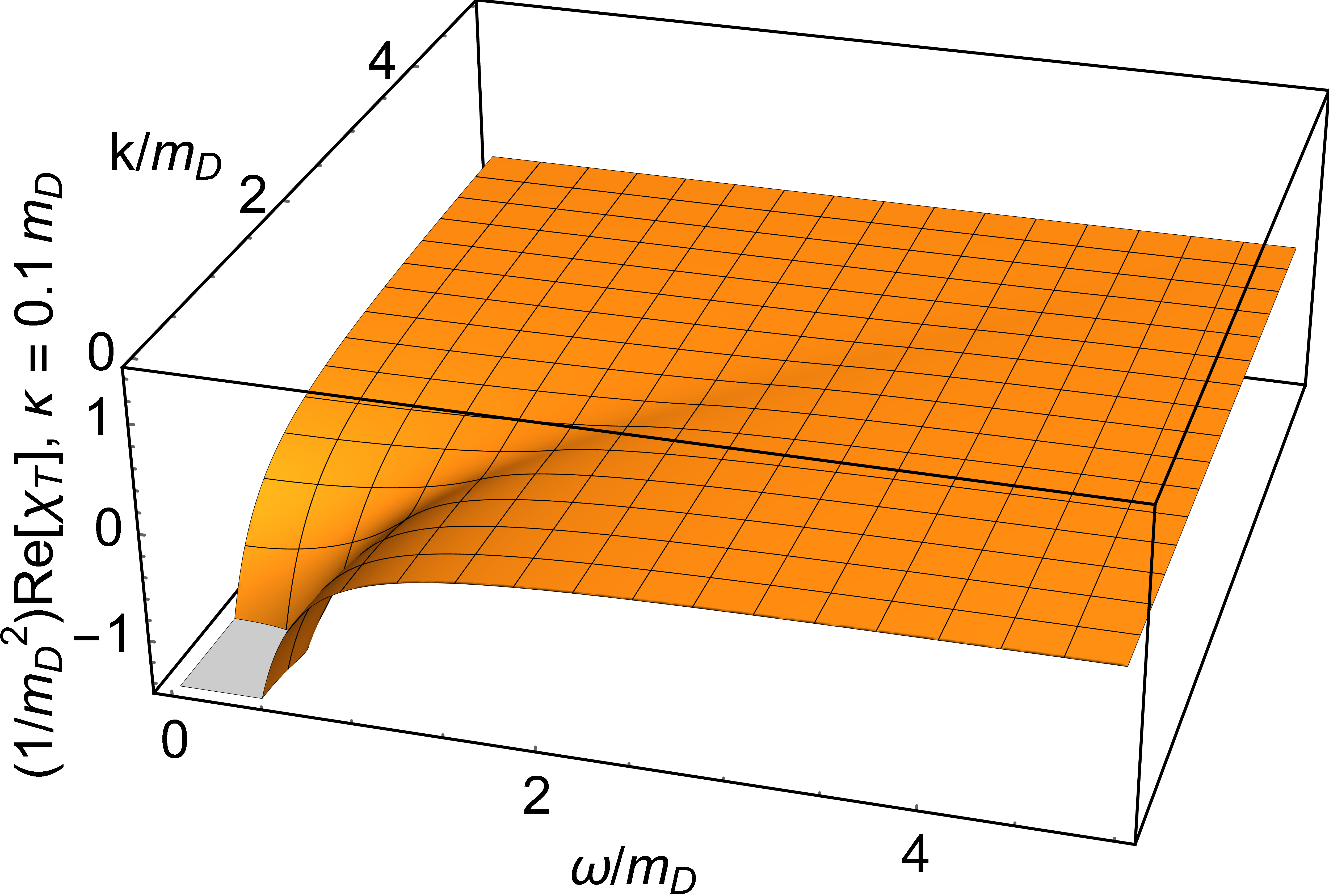}
\hspace{0.05\linewidth}
\includegraphics[width=0.45\linewidth]{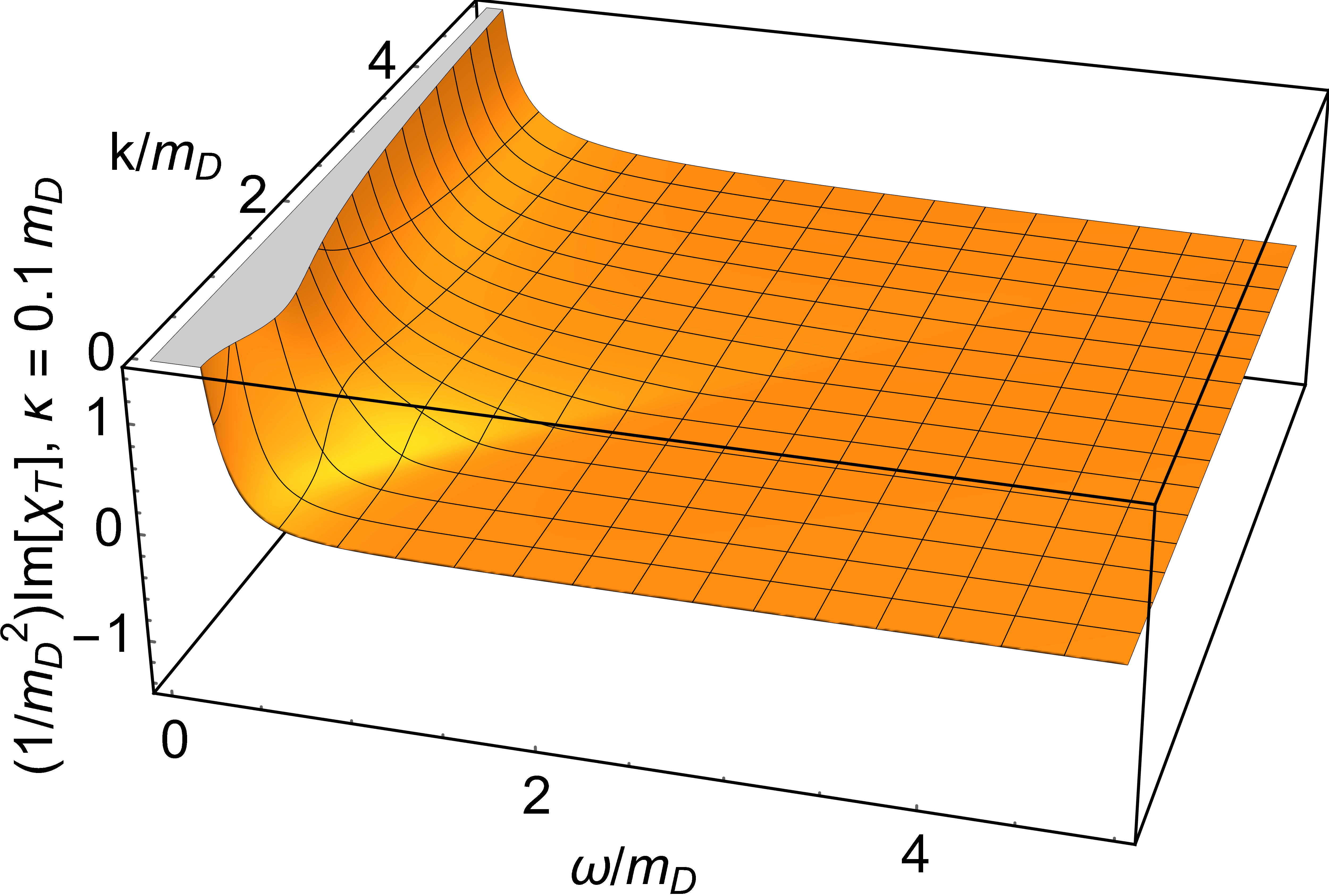}
\caption{Transverse susceptibility $\chi_T(\omega, k)$ in units of $m_D^2$ for $\kappa = 0.1\,m_D$. Left panel: real part of $\chi_T(\omega,k)$; right panel: imaginary part of $\chi_T(\omega,k)$. }
\label{fig:chi_T3d}
\end{figure}
\end{widetext}

We now compare our results for the BGK collision term with those obtained for the simple Anderson-Witting (AW) collision model (\ref{eq:lincoll}) which does not contain the current conserving BGK modification. This is of interest as such results appear frequently in literature. In the ultrarelativistic limit, the polarization tensor with the AW collision term is given by the tensor components $R^\mu_\nu$ from equations (\ref{eq:r00}-\ref{eq:rzz}) in Appendix \ref{sec:ultrarel}. Because $R^\mu_\nu$ alone does not satisfy the current conservation condition (\ref{eq:currentconservcomp}) there are two different possible definitions of longitudinal susceptibility $\chi_L$ constructed from either $R^z_z$ component (\ref{eq:rzz}) or the $R^0_0$ component (\ref{eq:r00}).

To avoid confusion, we continue to call the susceptibility constructed from the $R^z_z$ component of the polarization tensor the longitudinal susceptibility $\chi_L$ and the susceptibility constructed from the $R^0_0$ component of the polarization tensor the charge susceptibility $\chi_0$. Of course, when the collision term satisfies current conservation, $\chi_0 = \chi_L$, as it is indeed the case for the BGK modified collision term (\ref{eq:boltzmanncov}). Because the current conservation is violated for $R^\mu_\nu$ the two definitions of $\Pi_L$ (\ref{eq:piLT}) do not agree with each other. In other words the solution of Maxwell equations for the scalar potential or the longitudinal component of the vector potential are no longer related as they would be when $\partial_\mu j^\mu = 0$.  

\begin{figure}
	\centering
	\includegraphics[width=0.95\linewidth]{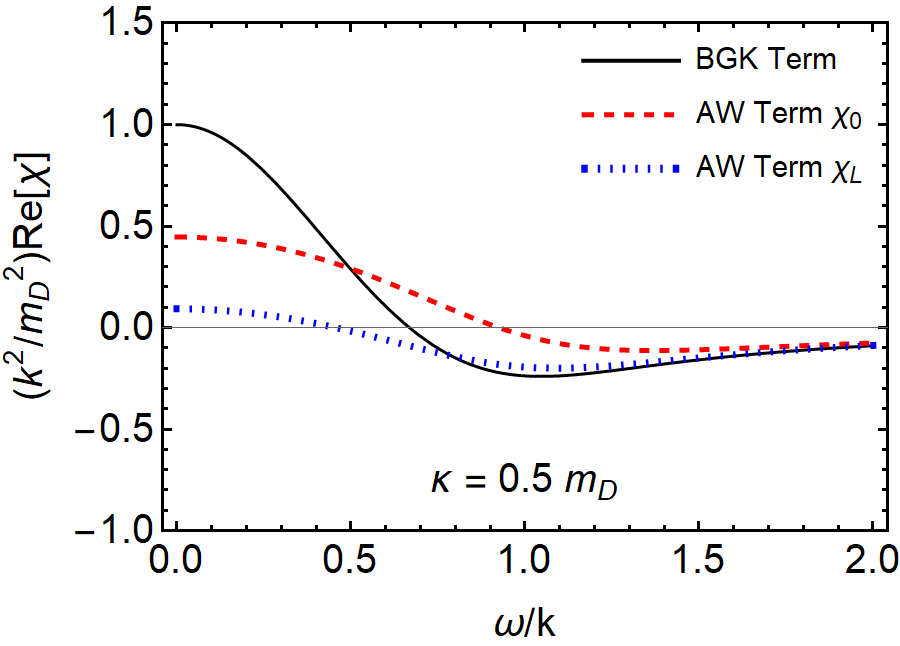} 
	\caption{Real part of $\chi_L$ as a function of $\omega/k$ for $\kappa = 0.5 m_D$. We compare the BGK collision model considered in this paper and Anderson-Witting (AW) model given by the collision term (\ref{eq:lincoll}) with both definitions of $\Pi_L$ (\ref{eq:piLT}).}	
	\label{fig:RechiLcomp}
\end{figure}

\begin{figure}
	\centering
	\includegraphics[width=0.95\linewidth]{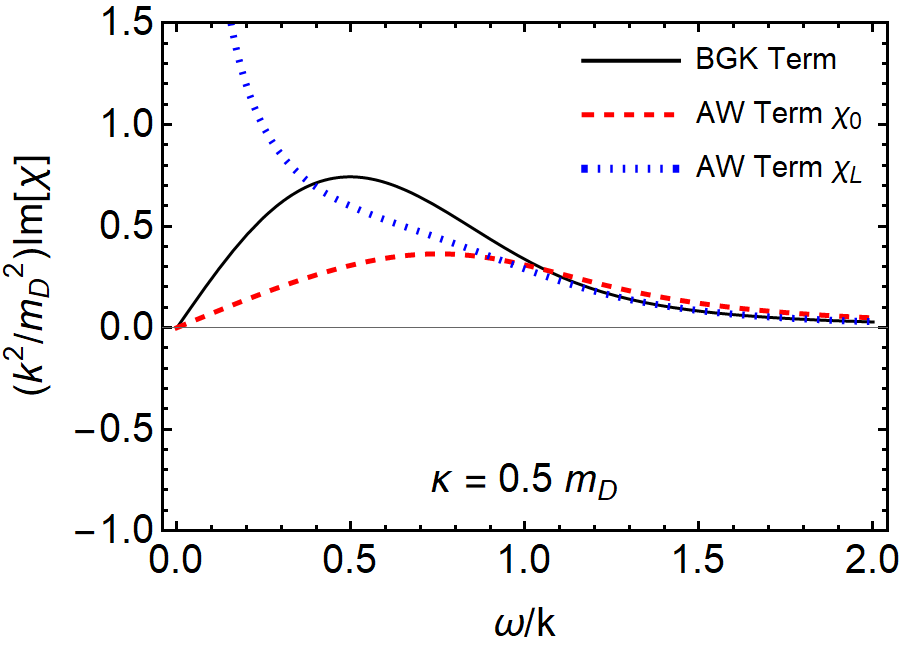}
	\caption{Imaginary part of $\chi_L$ as a function of $\omega/k$ for $\kappa = 0.5 m_D$. We compare the BGK collision model considered in this paper and Anderson-Witting (AW) model given by the collision term (\ref{eq:lincoll}) with both definitions of $\Pi_L$ (\ref{eq:piLT}).}
	\label{fig:ImchiLcomp}
\end{figure}

In Figs.~\ref{fig:RechiLcomp} and \ref{fig:ImchiLcomp} we compare the real and imaginary components of the longitudinal and charge susceptibility in the BKG and AW collision models for $\kappa = 0.5 m_D$. For the BGK model there is only a single curve (the solid lines); there are two different curves for the unmodified AW model: The dashed lines show the charge susceptibility $\chi_0$; the dotted lines show the longitudinal susceptibility $\chi_L$. The fact that the two lines differ, especially in the region $\omega < k$, is a direct consequence of current conservation violation of the AW model. Our results (solid curves) are clearly different from either of the AW curves.

We note in passing that the $\chi_L$ of the AW model, as defined by the $R^z_z$ component of the polarization tensor, is the same as we would get for the Blaizot-Iancu \cite{Blaizot:2001nr} form of the polarization tensor 
\begin{equation}
\Pi^\mu_\nu(k) = m_D^2 \left(-\delta^{\mu0}\delta_{\nu0} + \omega \int \frac{d\Omega}{4\pi}\frac{v^\mu v_\nu}{v \cdot k + i \kappa} \right)
\end{equation}
by simply substituting a nonzero value $\kappa$. However, note that this version suffers from the limitation that it is not manifestly Lorentz covariant because the 4-velocity of the medium $u^\mu$ is not explicitly introduced (compare with the covariant expression for $R^\mu_\nu$ (\ref{eq:Rmunu}) and its evaluation in Appendix \ref{sec:ultrarel}).

\subsection{Conductivity}
\begin{figure}
	\centering
	\includegraphics[width=0.95\linewidth]{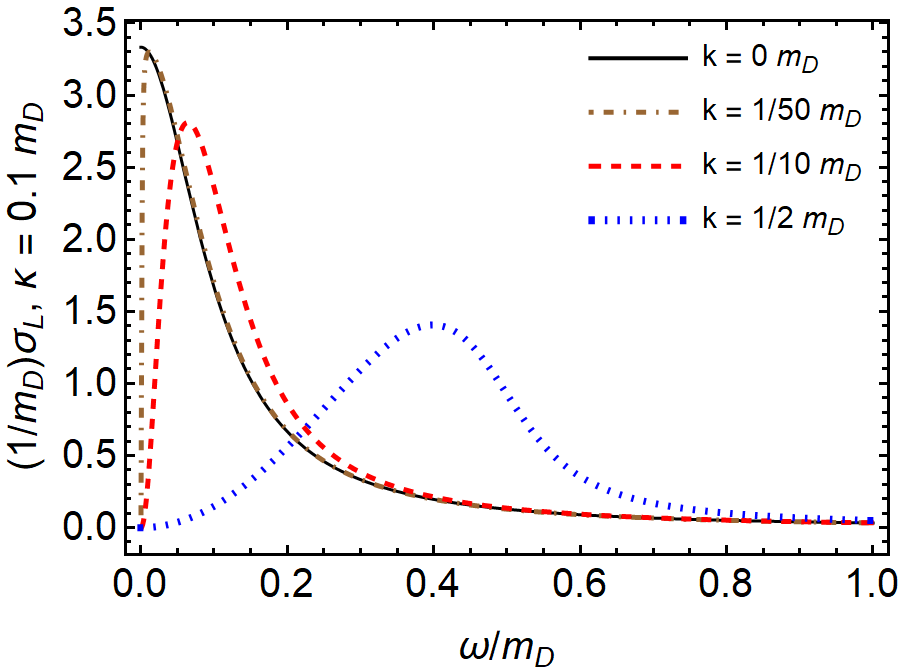} 
	\caption{Real part of $\sigma_L$ for different values of $k$.}	
	\label{fig:sigma_L}
\end{figure}

\begin{figure}
	\centering
	\includegraphics[width=0.95\linewidth]{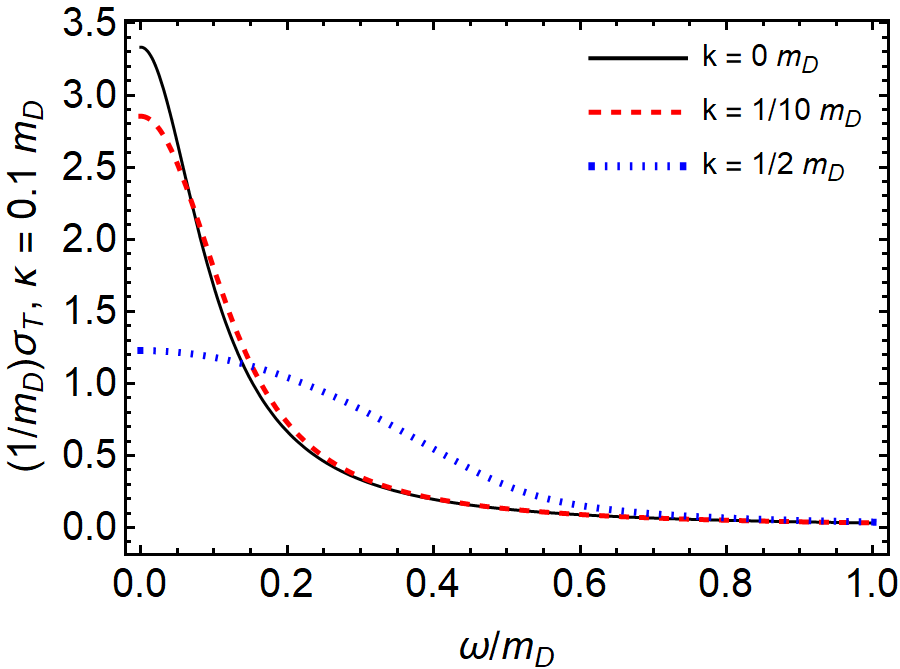}
	\caption{Real part of $\sigma_T$ for different values of $k$.}
	\label{fig:sigma_T}
\end{figure}
The conductivity tensor is commonly defined by the spatial part of the linear response equation (\ref{eq:ohm}), where the vector potential is expressed in terms of the electric field $i \omega \widetilde{A^i} = \widetilde{E^i}$. One finds Refs.~\cite{Starke:2014tfa,Melrose:2008}:
\begin{align}
    \sigma_T(\omega,k) &\equiv - i \omega \chi_T(\omega, k)\,,\\
    \sigma_L(\omega,k) &\equiv - i \omega \chi_L(\omega, k)\,.
\end{align}
We only consider the real part of the conductivity for the two independent response functions ($\Pi_L$) and ($\Pi_T$), which are shown in Figs.~\ref{fig:sigma_L} and \ref{fig:sigma_T} for three different values of $k$. 
Looking at the $k \to 0$ limit of the transverse susceptibility in equation (\ref{chi_l_opt}), we see that the transverse and longitudinal conductivities have the same Drude model dependence on frequency~\cite{Drude:1900} with $\tau = 1/\kappa$:
\begin{equation}
    \sigma_L(\omega,0) = \sigma_T(\omega,0) = \frac{\omega_p^2\tau}{1-i \omega\tau} \,,
\end{equation}
which reproduces Satow's gradient expansion~\cite{Satow:2014lia}. We note the discontinuous behavior of longitudinal conductivity
\begin{equation}
\lim_{k\rightarrow 0}\sigma_L(0,k) \neq \lim_{\omega\rightarrow 0}\sigma_L(\omega,0)\,,
\end{equation}
which originates in the infinite extent of the plasma considered in this work. For finite systems the longitudinal conductivity vanishes in the limit $\omega \rightarrow 0$ even in the $k = 0$ case \cite{Baranger:1989}.
\subsection{Dispersion Relations} \label{sec:disp}

The propagator of electromagnetic perturbations of the plasma is obtained by inverting Maxwell's equations including the induced current:
\begin{equation}
    -ik_{\mu}\widetilde{F}^{\mu \nu} = \mu_0( \widetilde{j}_{\mathrm{ind}}^{\nu}+\widetilde{j}_{\mathrm{ext}}^{\nu})
\end{equation}
Including the induced current on the left-hand side of the equation and writing the expression in-terms of $A^{\mu}$ one finds,
\begin{equation}
    (k^2g^{\mu \nu} - k^{\mu} k^{\nu} + \mu_0\Pi^{\mu \nu})\widetilde{A}_{\nu} = - \mu_0\widetilde{j}_{\mathrm{ext}}^{\nu} \,.
\end{equation}
The propagator $D^\mu_\nu(k)$ is obtained by inverting the equation:
\begin{equation}
    \widetilde{A}_{\nu}(k) = -D^{\mu}_{\nu}(k) \,\widetilde{j}_{\mathrm{ext}}^{\nu}(k) \,.
\end{equation}
For an isotropic medium the poles in the propagator can be expressed in terms of $\Pi_L$ and $\Pi_T$; the result is~\cite{Melrose:2008}:
\begin{equation}
\left[(k\cdot u)^2+ \mu_0\Pi_L(\omega, k)\right]\left[k^2 + \mu_0 \Pi_T(\omega, k)\right]^2=0 \,,
\end{equation}
where again $u$ is the 4-velocity of the medium rest frame. Note that the transverse mode has duplicate solutions as it describes modes in a plane perpendicular to $k$.

\begin{figure}
    \centering
    \includegraphics[width=0.95\linewidth]{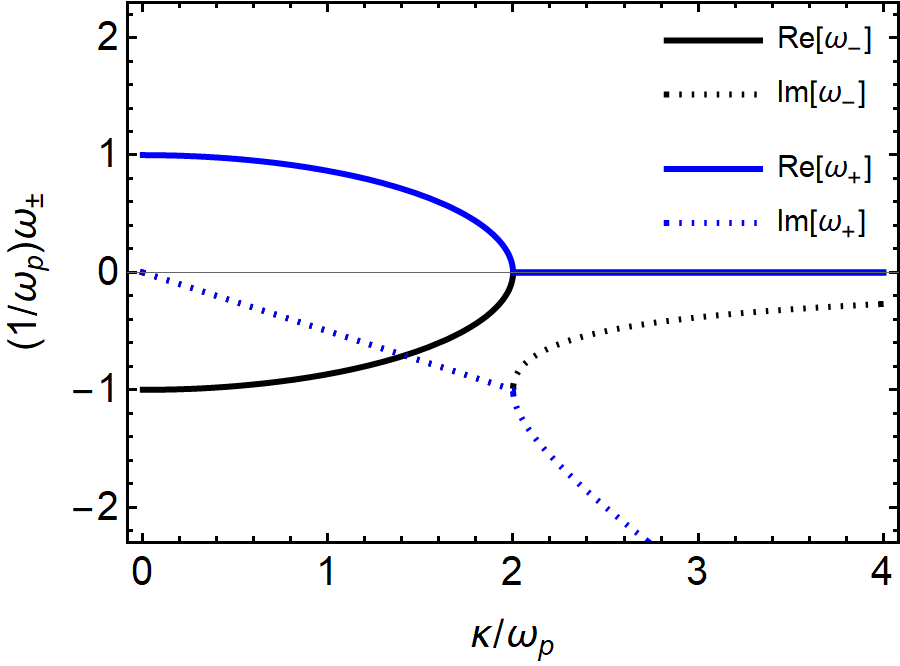}
    \caption{Real and imaginary part of the plasma frequencies $\omega_\pm$ as a function of $\kappa$. 
    The region $\kappa>2\omega_p$ corresponds to overdamping, with one mode becoming more quickly damped 
    and the other more slowly damped.}
    \label{fig:plasma-freq}
\end{figure}

In the limit of $k \to 0$ both the transverse and longitudinal roots of the dispersion relation reduce to the frequency of plasma oscillations (Figure \ref{fig:plasma-freq}):
\begin{equation}\label{plasmafreq}
    \omega_{\pm} = -\frac{i\kappa}{2} \pm \sqrt{\omega_p^2 - \frac{\kappa}{4}^2}\,,
\end{equation}
where the plasma frequency $\omega_p$ is explicitly given in the ultrarelativistic and non-relativistic limits, respectively, by:
\begin{equation}
\omega_p^2 = \frac{1}{3} m_D^2 \quad (\mathrm{UR})\,, \qquad \omega_p^2 = m_L^2 \quad (\mathrm{NR}) \,.
\end{equation}
This expression gives a physical meaning to the mass $m_L$ introduced in (\ref{eq:mL}).

For oscillatory waves of the form $E=E_0e^{-i\omega t}$, in the case of $\kappa \ll \omega_p$ the waves are weakly damped. For $\kappa > 2\omega_p$, the square root is imaginary and the wave becomes overdamped. For $\kappa \gg \omega_p$ the weakly damped solution has a long lifetime $\tau \approx \kappa/\omega_p^2$. The plasma frequency is plotted below as a function of $\kappa$.

In the static limit $\omega \rightarrow 0$ the solutions to the longitudinal root of the dispersion relation go to
\begin{equation}
    k = \pm  i m_D \,.
\end{equation}
The positive imaginary root describes the Debye screening of a stationary charge within the plasma. The analysis of the real and imaginary parts of the zeros for the $\kappa$-dependent dispersion relation for both transverse and longitudinal modes can be found in \cite{Carrington:2003je,Schenke:2006xu}. Our approach starting from manifestly covariant formulation with arbitrary medium 4-velocity matches these results.

\section{Summary and Conclusions}

The main result of this work is a manifestly Lorentz covariant description of the dynamics of perturbations in a relativistic plasma created by external or intrinsically generated electromagnetic fields in the presence of collisional damping within the range of conditions where linear plasma response applies. The covariant polarization tensor, which is gauge invariant and current conserving, agrees in the ultra-relativistic limit with earlier derivations in the plasma rest frame~\cite{Carrington:2003je,Schenke:2006xu}. In the limit of vanishing collisional damping ($\kappa \to 0$) our results reduce to the well known expressions originally obtained by Weldon~\cite{Weldon:1982aq}. For non-zero values of the collisional damping rate our results provide for an effective analytical description of the dynamics of relativistic plasmas under more realistic conditions, extending their applicability to a much wider range of phenomena.

Our original motivation for undertaking this work was
\begin{enumerate}
\item[1.]
The study of quark-gluon plasma response to the extreme electromagnetic fields produced in heavy ion collisions.  
\end{enumerate}
Of particular relevance here is that the results for electric conductivity (see Figures \ref{fig:sigma_T} and \ref{fig:sigma_L}) show strong frequency dependence governed by the average collision rate $\kappa$. 

Moreover, we anticipate that the approach developed here and its clarification of prior results will serve as a point of departure for the exploration and study of several other distinct domains of physics. Notable examples are:
\begin{enumerate}
\item[2.] Plasma physics characterizing the dynamics of cosmological evolution from the big-bang to the recombination era;
\item[3.] Stellar evolution from the formation of stars to terminal processes such as supernovae;
\item[4.] Plasma particle acceleration and/or high intensity laser interaction with matter generated in the laboratory in response to transient extreme electric and magnetic fields.
\end{enumerate}

For most of these applications, multiple plasma components, i.~e.\ heavy (ions) and light (electrons) charged particles, will need to be considered. For simplicity, we focused here on a one-component plasma. In many situations, chemical potentials are required to properly describe the asymmetry between particle and antiparticle distributions or to account for not fully saturated phase space distributions. Again for simplicity, here we assumed particle-antiparticle symmetry and fully saturated phase space. Also, in a multi-component plasma some plasma components may be better described in the nonrelativistic limit (see Appendix~\ref{sec:nonrel}) while others may remain relativistic as studied in the main body of this work.

As we already mentioned, our treatment can be generalized and extended in multiple ways, which generally require a modest amount of numerical computation. Specific examples of such generalizations are: a 4-momentum conserving extension of the collision term; a momentum-dependent collisional relaxation rate $\kappa(p)$; plasma kinetics that is neither nonrelativistic nor ultrarelativistic; degenerate or undersaturated phase-space populations. We intend to pursue these extensions and applications in future publications.

\begin{acknowledgments}
The research of MF and CMG was supported by the US Department of Energy under Grant Contract DE-SC0012704 to the Brookhaven National Laboratory. BM acknowledges support by the U.S. Department of Energy Office of Science under Grant DE-FG02-05ER41367. We thank Harold Baranger and Stanis\l aw Mr\'owczy\'nski for useful discussions regarding interpretations of our results and the referee for making us aware of the work of Schenke {\it et al.}.
\end{acknowledgments}

\appendix
\section{Ultrarelativistic limit}\label{sec:ultrarel}

For the calculation of the components of the polarization tensor (\ref{eq:pimunu}) we consider the rest frame of the medium $u^\mu = (c,0)$ and the massless limit of particles moving at the speed of light
\begin{align}
\label{eq:massless1}p^\mu &= |\pmb{p}|v^\mu,\\
 v^\mu &= (1, \sin\theta\cos\varphi, \sin\theta\sin\varphi,\cos\theta)\,.
\end{align}
That way the product $p\cdot u$ can be expressed as
\begin{equation}
\label{eq:massless2}p\cdot u = p^0 = |\pmb{p}|\,.
\end{equation}
Without loss of generality we choose to orient the $\pmb{k}$ vector along the $z$-axis. That way the covariant $k^\mu$ reads
\begin{equation}
\label{eq:kdirect}k^\mu = (\omega, 0,0,k)\,.
\end{equation}
the product $p \cdot k$ is given by
\begin{equation}
p \cdot k = |\pmb{p}|(\omega - k \cos \theta)\,.
\end{equation}
For example the $n_\mathrm{eq}$ integral (\ref{eq:ndef2}) in the approximation of mass-less particles
\begin{equation}
n_\mathrm{eq} = \frac{1}{\pi^2}\int_0^\infty |\pmb{p}|^2d|\pmb{p}| f_\mathrm{eq}(|\pmb{p}|) = \frac{3T^3}{2\pi^2}\zeta(3)\,, 
\end{equation}
because the angular integration in the rest frame of the medium is trivial.
\subsection{Evaluating $R^\mu_\nu(k)$}
The definition of $R^\mu_\nu(k)$ is given in (\ref{eq:Rmunu}). Under the conditions described above (\ref{eq:massless1}-\ref{eq:kdirect}) we have
\begin{multline}
	R^\mu_\nu = -2q^2\int_0^\infty \frac{|\pmb{p}|^2d|\pmb{p}|}{\pi^2 }f'_\mathrm{eq}(|\pmb{p}|)\\
	\times \int \frac{d\Omega}{4\pi}\frac{\omega v^\mu v_\nu - (\omega - k \cos \theta)v^\mu \delta_{\nu 0}}{\omega - k \cos \theta + i\kappa}\,.
\end{multline}
We see that the integrals over the magnitude of momentum and the angular integrals in this limit separated. The integral over the magnitude is customarily called Debye screening mass
\begin{equation}\label{eq:mD}
m_D^2 \equiv - \frac{2q^2}{\pi^2}\int_0^\infty |\pmb{p}|^2d|\pmb{p}| f'_\mathrm{eq}(|\pmb{p}|) = \frac{q^2T^2}{3}\,.
\end{equation}
The angular integration eliminates most of the tensor components. The only non-zero terms are
\begin{align}
	\label{eq:r00}R^0_0 &= - m_D^2 L\,,\\
	R^x_x = R^y_y &= \frac{m_D^2\omega}{4k} \left(\frac{\omega'^2}{k^2}\Lambda - \Lambda - \frac{2\omega'}{k}\right)\,,\\
	\label{eq:r0z}R^0_z &=  m_D^2\frac{\omega}{k}L\,,\\
	R^z_0 & = - m_D^2\frac{\omega'}{k}L\,,\\
	\label{eq:rzz}R^z_z &= m_D^2\frac{\omega \omega'}{k^2}L\,,
\end{align}
where the quantities $\omega'$, $\Lambda$, and $L$ are defined in (\ref{eq:definitions}).

\subsection{Evaluating $Q(k)$}

The scalar term $Q(k)$ (\ref{eq:Q}) can be also split into integration over the magnitude of momentum and angular part under conditions (\ref{eq:massless1}-\ref{eq:kdirect})
\begin{multline}
	Q(k) = - i \frac{\kappa}{n_\mathrm{eq}} \int_0^\infty \frac{|\pmb{p}|^2d|\pmb{p}|}{\pi^2}f_\mathrm{eq}(|\pmb{p}|)\\
	\times\frac{1}{4\pi}\int\frac{d\Omega}{\omega - k\cos\theta + i\kappa}\,. 
\end{multline}
The integral over the magnitude of $|\pmb{p}|$ reads
\begin{equation}
- \frac{i}{\pi^2} \frac{\kappa}{n_\mathrm{eq}} \int_0^\infty |\pmb{p}|^2d|\pmb{p}|f_\mathrm{eq}(|\pmb{p}|) = -i\kappa\,,
\end{equation}
where the dependence on $n_\mathrm{eq}$ exactly cancelled. The angular integral is
\begin{equation}
\frac{1}{4\pi}\int\frac{d\Omega}{\omega - k\cos\theta + i\kappa} = \frac{1}{2k}\Lambda\,.
\end{equation}
Altogether we have for $Q(k)$
\begin{equation}\label{eq:qres}
\boxed{Q(k) = -\frac{i\kappa}{2k}\Lambda\,.}
\end{equation}

\subsection{Evaluating $Q^\mu(k)$}

Repeating the same logic for the $Q^\mu(k)$ term (\ref{eq:Qmu}) we have
\begin{multline}
	Q^\mu(k) = - 2qi \frac{\kappa}{n_\mathrm{eq}} \int_0^\infty \frac{|\pmb{p}|^2d|\pmb{p}|}{\pi^2} f_\mathrm{eq}(|\pmb{p}|)\\
	\times \frac{1}{4\pi}\int \frac{v^\mu d\Omega}{\omega - k \cos \theta + i\kappa}\,.
\end{multline}
The integral over the magnitude $|\pmb{p}|$ is
\begin{equation}
- \frac{2qi}{\pi^2} \frac{\kappa}{n_\mathrm{eq}} \int_0^\infty |\pmb{p}|^2d|\pmb{p}|f_\mathrm{eq}(|\pmb{p}|) = - 2iq\kappa\,,
\end{equation}
where again the $\zeta(3)T^3$ dependence of $n_\mathrm{eq}$ cancelled. The nonzero contributions of the angular part are
\begin{align}
	I^0 = \frac{1}{4\pi}\int \frac{d\Omega}{\omega - k\cos\theta + i\kappa} &= \frac{1}{2k}\Lambda\,,\\
	I^z = \frac{1}{4\pi}\int \frac{\cos \theta d\Omega}{\omega - k\cos\theta + i\kappa}& = - \frac{1}{k}L\,.
\end{align}
Therefore the non-zero components of $Q^\mu(k)$ are
\begin{equation}\label{eq:qmures}
\boxed{Q^0 =- \frac{iq\kappa}{k}\Lambda, \quad Q^z = \frac{2iq\kappa}{k}L\,.}
\end{equation}
\subsection{Evaluating $H_\nu(k)$}

Separating the itegration into the angular and magnitude part from the definition of $H_\nu$ (\ref{eq:Hnu}) we have
\begin{multline}
	H_\nu(k) = - q \int_0^\infty \frac{|\pmb{p}|^2 d|\pmb{p}|}{\pi^2}f'_\mathrm{eq}(|\pmb{p}|)\\
	\times \int \frac{d\Omega}{4\pi} \frac{\omega v_\nu - (\omega - k\cos\theta)\delta_{\nu 0}}{\omega - k \cos \theta + i\kappa}\,.
\end{multline}
The integral over magnitude $|\pmb{p}|$ is
\begin{equation}
-\frac{q}{\pi^2}\int_0^\infty |\pmb{p}|^2d|\pmb{p}|f'_\mathrm{eq}(|\pmb{p}|)= \frac{qT^2}{6}
\end{equation}
and the angular integral has only two non-zero components
\begin{align}
	I_0 &= \int \frac{d\Omega}{4\pi} \frac{k\cos\theta}{\omega-k\cos\theta+i\kappa}= - L\,,\\
	I_z &= -\int \frac{d\Omega}{4\pi} \frac{\omega \cos \theta}{\omega - k\cos \theta + i \kappa} = \frac{\omega}{k}L\,.
\end{align}
The change in sign is due to $v_z = - \cos\theta$ because of the lowered index. Altogether the non-zero components of $H_\nu(k)$ are
\begin{equation}\label{eq:hnures}
\boxed{H_0 = -\frac{qT^2}{6}L, \quad H_z = \frac{qT^2 \omega}{6 k}L\,.}
\end{equation}

\subsection{Summary of the components}

Using the results (\ref{eq:r00}-\ref{eq:rzz}) and (\ref{eq:qres},\ref{eq:qmures},\ref{eq:hnures}) the components of the polarization tensor (\ref{eq:pimunu}) are
\begin{align}
	\label{eq:pi00}\Pi^0_0  &= - m_D^2 L \left( 1+ \frac{i\kappa\Lambda}{2k-i\kappa\Lambda} \right)\,,\\
	\label{eq:pi0z}\Pi^0_z &=m_D^2 \frac{\omega}{k}L \left( 1+ \frac{i\kappa\Lambda}{2k-i\kappa\Lambda} \right)\,,\\
	\label{eq:piz0}\Pi^z_0 &= - m_D^2 L \left( \frac{\omega'}{k} - \frac{2i\kappa L}{2k-i\kappa\Lambda} \right)\,,\\
	\label{eq:pizz}\Pi^z_z &= m_D^2 \frac{\omega}{k}L \left( \frac{\omega'}{k} - \frac{2i\kappa L}{2k-i\kappa\Lambda} \right)\,.
\end{align}
Additionally the transversal components are given just by $R^\mu_\nu$
\begin{equation}
\Pi^x_x = \Pi^y_y = \frac{m_D^2\omega}{4k}\left( \frac{\omega'^2}{k^2}\Lambda - \Lambda - \frac{2\omega'}{k}\right)\,.
\end{equation} 
The gauge invariance conditions (\ref{eq:gaugeincomponents}) are obviously satisfied. In order to prove the current conservation condition (\ref{eq:currentconservcomp}) we need to combine the $i\kappa$ proportional term from $\omega'$ in the first term in (\ref{eq:piz0}) with the second term and recover the second term in (\ref{eq:pi00}) times $\omega/k$
\begin{multline}
	\frac{i\kappa}{k} - \frac{2i\kappa L}{2k - i\kappa \Lambda} = \frac{\kappa^2 \Lambda + i \kappa \omega' \Lambda}{k(2k-i\kappa \Lambda)} =\frac{\omega}{k} \frac{i\kappa \Lambda}{2k-i\kappa \Lambda)}\,.
\end{multline}
In the first step we substituted the definition of $L$ (\ref{eq:definitions}) and in the second step the $i\kappa$ part of $\omega'$ in the numerator cancels and we obtain $\omega/k$ times the second term of (\ref{eq:pi00}) as expected. This also means that the last two components of $\Pi^\mu_\nu$ (\ref{eq:piz0},\ref{eq:pi0z}) can be written equivalently as
\begin{align}
	\Pi^z_0 &= -m_D^2\frac{\omega}{k}L\left(1 + \frac{i\kappa \Lambda}{2k - i\kappa \Lambda}\right)\,,\\
	\Pi^z_z &= m_D^2\frac{\omega^2}{k^2}L\left(1 + \frac{i\kappa \Lambda}{2k - i\kappa \Lambda}\right)\,.
\end{align}
In this form we see explicitly that baring the lowering of the spatial index
\begin{equation}\label{eq:symmetry}
\Pi^0_z = -\Pi^z_0\,.
\end{equation}
the tensor $\Pi_{\mu\nu}$ is symmetric. 

\section{Non-relativistic limit}\label{sec:nonrel}

We now consider the low-temperature non-relativistic limit. For low temperatures ($T \ll m$) the Fermi distribution can be replaced by the Boltzmann factor
\begin{equation}
f_\mathrm{eq} \approx \exp\left(- \frac{p \cdot u}{T}\right)
\end{equation}
and in the non-relativistic limit we approximate the energy in the rest frame by its nonrelativistic expansion
\begin{equation}
(p \cdot u) = p^0 = m \left(1 + \frac{|\pmb{p}|^2}{2m^2} + \ldots \right)\,.
\end{equation}
In the following we will keep only the first term of this expansion. For example we can evaluate the $n_\mathrm{eq}$ integral (\ref{eq:ndef2}) as
\begin{equation}
n_{eq} = \int \frac{d\Omega}{4\pi}\frac{|\pmb{p}|^2d|\pmb{p}|}{\pi^2}\exp\left[-\frac{m}{T}\left(1 + \frac{1}{2}\frac{|\pmb{p}|^2}{m^2}\right)\right]\,.
\end{equation}
The angular integration is trivial and the rest gives us a gaussian integral
\begin{equation}
n_\mathrm{eq} = \frac{1}{4}e^{-m/T}\left(\frac{2mT}{\pi} \right)^{3/2}\,.
\end{equation}
The denominator of all the integrals reads
\begin{equation}
p \cdot k + i(p\cdot u)\kappa = p^0\left(\omega - \frac{|\pmb{p}|}{p^0}k\cos \theta +i\kappa\right)
\end{equation}
and in the first order 
\begin{equation}
\frac{|\pmb{p}|}{p^0} \approx \frac{|\pmb{p}|}{m}\,.
\end{equation}
Unfortunately, as we will see below, the angular and magnitude integration no longer factorize. We will resolve this problem by computing all angular integrals up to the quadratic order in the small parameter $|\pmb{p}|/m$.

\subsection{Evaluation of $R^\mu_\nu(k)$}

The definition of $R^\mu_\nu(k)$ is given in (\ref{eq:Rmunu}). In non-relativistic and low-temperature limit we have
\begin{multline}
	R^\mu_\nu(k) = -2q^2 \int_0^\infty \frac{|\pmb{p}|^2 d|\pmb{p}|}{\pi^2}f'_\mathrm{eq}(|\pmb{p}|)\\
	\times \frac{1}{p^0}\int \frac{d\Omega}{4\pi} \frac{\omega p^\mu p_\nu - (k \cdot p)p^\mu \delta_{0\nu}}{k \cdot p+ i(p \cdot u)\kappa}\,.
\end{multline}
Introducing the abbreviation
\begin{equation}
g(\cos\theta) = \omega -|\pmb{p}|k\cos\theta / m + i\kappa \,,
\label{eq:dcostheta}
\end{equation}
the angular integrals are easily evaluated in up to quadratic order of $|\pmb{p}|/m$:
\begin{align}
	I^0_0 &= \int \frac{d\Omega}{4\pi} \frac{|\pmb{p}|k\cos\theta}{g(\cos\theta)m} \approx
	\frac{1}{3}\left(\frac{|\pmb{p}|k}{m\omega'}\right)^2\,,\\
	I^0_z &= - \int \frac{d\Omega}{4\pi} \frac{\omega |\pmb{p}|\cos\theta}{g(\cos\theta)m} \approx - \frac{1}{3}\frac{\omega|\pmb{p}|^2k}{\omega'^2m^2}\,,\\
	I^z_0 &= \int \frac{d\Omega}{4\pi}\frac{|\pmb{p}|^2 k \cos^2\theta}{g(\cos\theta)m^2} \approx \frac{1}{3} \frac{|\pmb{p}|^2 k}{\omega' m^2}\,,\\
	I^z_z &= - \int \frac{d\Omega}{4\pi} \frac{|\pmb{p}|^2 \omega \cos^2\theta}{g(\cos\theta)m^2} \approx - \frac{1}{3}\frac{|\pmb{p}|^2\omega}{m^2 \omega'}\,,\\
	I^x_x &= I^y_y = - \int \frac{d\Omega}{4\pi} \frac{\omega |\pmb{p}|^2 \sin^2\theta \cos^2 \varphi}{g(\cos\theta)m^2} \approx - \frac{1}{3}\frac{|\pmb{p}|^2\omega}{m^2 \omega'}\,.
\end{align}
In order for the expansion to be possible $|\pmb{p}|k/m\omega' \ll 1$ which places also limit on $k$ and $\omega'$. All the magnitude integrals are now of the type
\begin{multline}\label{eq:mL}
	m_L^2 \equiv \frac{2q^2}{3m^2\pi^2}\int_0^\infty |\pmb{p}|^4 d|\pmb{p}| \left(\frac{1}{T} e^{-p^0/T} \right)\\
	= q^2 \left(\frac{2mT}{\pi}\right)^{3/2}\frac{e^{-m/T}}{2m}\,,
\end{multline}
where $m_L^2$ defines a mass scale. Altogether
\begin{equation}\label{eq:r00nonrel}
R^0_0 = m_L^2 \frac{k^2}{\omega'^2}\,, \quad R^0_z = -m_L^2 \frac{\omega k}{\omega'^2}\,,
\end{equation}
\begin{equation}\label{eq:rz0nonrel}
R^z_0 = m_L^2 \frac{k}{\omega'}\,, \quad R^x_x = R^y_y = R^z_z = -m_L^2 \frac{\omega}{\omega'}\,.
\end{equation}

\subsection{Evaluation of $Q(k)$}

In the low temperature and non-relativistic limit the integral for $Q(k)$ (\ref{eq:Q}) becomes
\begin{multline}
	Q(k) = -i \frac{\kappa}{n_\mathrm{eq}} \int_0^\infty \frac{|\pmb{p}|^2d|\pmb{p}|}{\pi^2} f_\mathrm{eq}(|\pmb{p}|) \\
	\times \int \frac{d\Omega}{4\pi} \frac{1}{\omega - |\pmb{p}|k \cos \theta/m +i\kappa}\,,
\end{multline}
in which again the integrals over angles and magnitude of the momentum $\pmb{p}$ not quite separate. Up to quadratic terms in $|\pmb{p}|/m$ the angular integral is
\begin{equation}
\int \frac{d\Omega}{4\pi} \frac{1}{g(\cos\theta)} \approx \frac{1}{\omega'}\left(1 + \frac{1}{3}\frac{|\pmb{p}|^2k^2}{m^2 \omega'^2}\right)\,,
\end{equation}
giving the result:
\begin{equation}\label{eq:qnonrel}
\boxed{Q(k) = -\frac{i\kappa}{\omega'} \left(1 + \frac{T k^2}{m\omega'^2} \right)\,.}
\end{equation}

\subsection{Evaluation of $Q^\mu(k)$}

In the low temperature and non-relativistic limit the integral for $Q^\mu(k)$ (\ref{eq:Qmu}) becomes
\begin{multline}
	Q^\mu(k) = -2qi \frac{\kappa}{n_\mathrm{eq}} \int_0^\infty \frac{|\pmb{p}|^2d|\pmb{p}|}{\pi^2} f_\mathrm{eq}(|\pmb{p}|)\\	
	\times \int \frac{d\Omega}{4\pi} \frac{p^\mu/m}{\omega - |\pmb{p}|k\cos\theta / m + i \kappa}\,,
\end{multline}
where the only non-zero angular integrals are up to second order in $|\pmb{p}|/k$
\begin{align}
	I^0 &= \int \frac{d\Omega}{4\pi} \frac{1}{g(\cos\theta)} \approx \frac{1}{\omega'} + \frac{1}{3} \frac{|\pmb{p}|^2 k^2}{m^2 \omega'^3}\,,\\
	I^z &= \int \frac{d\Omega}{4\pi} \frac{|\pmb{p}|\cos\theta}{g(\cos\theta)m} \approx \frac{1}{3} \frac{|\pmb{p}|^2k}{m^2 \omega'^2}\,.
\end{align}
Finally, after integration over the magnitude we have for the two non-zero components of $Q^\mu(k)$
\begin{equation}\label{eq:qmunonrel}
\boxed{Q^0 = - \frac{2qi\kappa}{\omega'}\left(1 + \frac{T k^2}{m\omega'^2} \right)\,, \quad Q^z = -2qi\kappa \frac{T k}{m\omega'^2}\,.}
\end{equation}

\subsection{Evaluation of $H_\nu(k)$}

In the low temperature and non-relativistic limit the integral for $H_\nu(k)$ (\ref{eq:Hnu}) becomes
\begin{multline}
	H_\nu(k) = - q \int_0^\infty \frac{|\pmb{p}|^2d|\pmb{p}|}{\pi^2} f'_\mathrm{eq}(|\pmb{p}|) \\
	\times \frac{1}{p^0} \int \frac{d\Omega}{4\pi} \frac{\omega p_\nu - (k \cdot p) \delta_{0\nu}}{\omega - |\pmb{p}|k\cos\theta / m + i\kappa}\,.
\end{multline}
The only non-zero angular integrals up to the second order in $|\pmb{p}|/k$ read
\begin{align}
	I_0 &= \int \frac{d\Omega}{4\pi} \frac{|\pmb{p}|k\cos\theta}{g(\cos\theta)m} \approx \frac{1}{3}\left(\frac{|\pmb{p}|k}{m\omega'}\right)^2\,,\\
	I_z &= - \int \frac{d\Omega}{4\pi} \frac{\omega|\pmb{p}|\cos\theta}{g(\cos\theta)m} \approx - \frac{1}{3} \frac{\omega |\pmb{p}|^2 k}{m^2 \omega'^2}\,.
\end{align}
After integration over $|\pmb{p}|$ we have two non-zero components of $H_\nu(k)$
\begin{equation}\label{eq:hnunonrel}
\boxed{H_0 = \frac{1}{2q} m_L^2 \frac{k^2}{\omega'^2}\,, \quad H_z = -\frac{1}{2q} m_L^2 \frac{\omega k}{\omega'^2}\,.}
\end{equation}

\subsection{Polarization tensor}

If we substitute the results obtained in the non-relativistic and low temperature limit (\ref{eq:r00nonrel},\ref{eq:rz0nonrel},\ref{eq:qnonrel},\ref{eq:qmunonrel},\ref{eq:hnunonrel}) back into (\ref{eq:pimunu}) we obtain the following expressions for the components of the polarization tensor
\begin{align}
	\Pi^0_0 &= m_L^2 \frac{k^2}{\omega'^2} \frac{1}{1-\frac{i\kappa}{\omega'}\left(1+\frac{T k^2}{m\omega'^2} \right)}\,,\\
	\label{eq:pi0znonrel}\Pi^0_z &= -m_L^2 \frac{\omega k}{\omega'^2} \frac{1}{1-\frac{i\kappa}{\omega'}\left(1+\frac{T k^2}{m\omega'^2} \right)}\,,\\
	\label{eq:piz0nonrel}\Pi^z_0 &= m_L^2 \frac{k^2}{\omega'^2} \left[\frac{\omega'}{k} + \frac{i\kappa\frac{T k}{m\omega'^2}}{1-\frac{i\kappa}{\omega'}\left(1+\frac{T k^2}{m\omega'^2} \right)} \right]\,,\\
	\label{eq:pizznonrel}\Pi^z_z &= -m_L^2 \frac{\omega k}{\omega'^2} \left[\frac{\omega'}{k} + \frac{i\kappa\frac{T k}{m\omega'^2}}{1-\frac{i\kappa}{\omega'}\left(1+\frac{T k^2}{m\omega'^2} \right)} \right]\,,\\	
	\Pi^x_x &= \Pi^y_y = -m_L^2 \frac{\omega}{\omega'}\,.	
\end{align}
	
The gauge conditions for the components (\ref{eq:gaugeincomponents}) are again obviously satisfied. The current conservation condition in components (\ref{eq:currentconservcomp}) requires further attention. Let's put the term in the parentheses of (\ref{eq:piz0nonrel}) on a common denominator and evaluate the numerator
\begin{multline}
	\frac{\omega'}{k}\left[1-\frac{i\kappa}{\omega'}\left(1+\frac{T k^2}{m\omega'^2} \right)\right] + i \kappa \frac{T k}{m\omega'^2}\\
	= \frac{\omega}{k} + \frac{i\kappa}{k}\left(1 + \frac{T k^2}{m\omega'^2}\right) - \frac{i\kappa}{k}\left(1 + \frac{T k^2}{m\omega'^2}\right) = \frac{\omega}{k}\,,
\end{multline}
as we were supposed to get. Therefore (\ref{eq:piz0nonrel}) and (\ref{eq:pizznonrel}) can be equivalently rewritten as
\begin{align}
	\Pi^z_0 &= m_L^2 \frac{\omega k}{\omega'^2} \frac{1}{1-\frac{i\kappa}{\omega'}\left(1+\frac{T k^2}{m\omega'^2} \right)}\,,\\	
	\Pi^z_z &= -m_L^2 \frac{\omega^2}{\omega'^2} \frac{1}{1-\frac{i\kappa}{\omega'}\left(1+\frac{T k^2}{m\omega'^2} \right)}\,.
\end{align}

\vspace*{0.1in}


\end{document}